\documentclass[a4paper,11pt]{article}

\usepackage{jinstpub} 

\usepackage{mediabb}
\usepackage{color}

\title{Angular dependence of columnar recombination in high pressure xenon gas using time profile of scintillation emission}

\author[a]{K.D.Nakamura,\note{Corresponding author.}}
\author[b]{S.Ban,}
\author[b]{M.Hirose,}
\author[b]{A.K.Ichikawa,}
\author[b]{Y.Ishiyama,}
\author[b]{A.Minamino,}
\author[a]{K.Miuchi,}
\author[b]{T.Nakaya,}
\author[c]{H.Sekiya,} 
\author[b]{S.Tanaka,}
\author[d]{K.Ueshima,}
\affiliation[a]{Kobe University,\\Rokkodai, Nada-ku Kobe-shi, Hyogo, 657-8501, Japan}
\affiliation[b]{Kyoto University,\\Kitashirakawaoiwake-cho Sakyo-ku Kyoto-shi Kyoto, 606-8502, Japan}
\affiliation[c]{Kamioka Observatory, ICRR, The University of Tokyo,\\456 Higashimozumi Kamioka-cho Hida-shi Gifu, 506-1205, Japan}
\affiliation[d]{RCNS, Tohoku University,\\6-3 Aramakiazaaoba, Aoba-ku Sendai-shi, Miyagi, 980-8578, Japan}

\emailAdd{kiseki@harbor.kobe-u.ac.jp}

\abstract{
	The angular dependence of the columnar recombination in xenon gas, if observed for low energy nuclear tracks, can be used for a direction-sensitive dark matter search.
	We measured both scintillation and ionization to study columnar recombination for 5.4 MeV alpha particles in a high pressure gas detector filled with 8 atm xenon.
	Since the recombination photons are emitted several~$\mu$s after de-excitation emission, scintillation photons are separated to the fast and slow components.
	The fast component does not show dependence on the track angle relative to the drift electric field, on the other hand, the slow component increases when the track is aligned with the electric field.
	The result indicates that the track angle relative to the electric field can be reconstructed from the scintillation time profile.
}


\begin{document}
	\maketitle
	\flushbottom

\section{Introduction}
A possibility of application of columnar recombination in high-pressure xenon gas to the direction-sensitive dark matter search was raised by D. R. Nygren in 2013 \cite{ref:NEXT_Nygren2013}.
Columnar recombination is a phenomenon in which recombination of electrons and ions distributed in a columnar form along the path of a charged particle increases when the electric field and the column are parallel in a time projection chamber.
With Xe of 0.05 g/cm$^3$ ($\sim$ 10 atm), the Onsager radius is 70 nm and the track length of recoil Xe of 30 keV is 2100 nm, a large aspect ratio of 30 times is expected \cite{ref:NEXT_Nygren2013}.
If the columnar recombination signal is significant and measurable, the angle with respect to the applied electric field can be reconstructed, for example, by comparing yields of scintillation and ionization.
A direction-sensitive dark matter search with a large target mass and spin-independent sensitivity would be possible if it is realized with high-pressure xenon gas.
In the previous study by the NEXT group, measurements were carried out by mixing TMA gas to xenon to quickly thermalize ionization electrons and also to enhance the Penning effect.
A correlation between the measured charge and the angle of $\alpha$-particle tracks was observed \cite{ref:NEXT_Herrera2014} but scintillation emission was strongly suppressed \cite{ref:NEXT_Nakajima2015}, so the conclusion was that it is difficult to apply it to a direction-sensitive dark matter search.
We examine to use the time profile of the scintillation from pure xenon gas to enhance the sensitivity to columnar recombination.
Recombinations and resulting photon emissions would happen during the drift of electrons.
Hence it is expected that its time scale is several $\mu$s and much slower than the de-excitation emission.
We conducted a measurement to quantify this effect with a small detector and measured the dependence on the track angle using 5.4 MeV $\alpha$-particle tracks.

\begin{figure}[t]
	\centering
	\includegraphics[width=120mm]{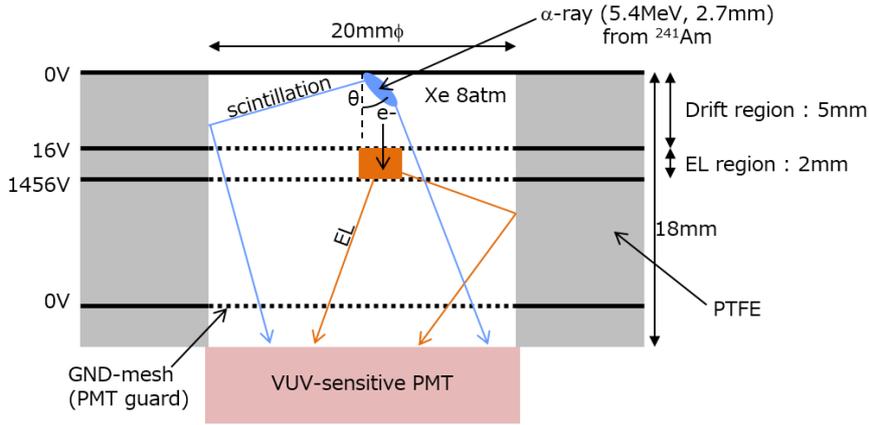}
	\caption{Schematic image of the detector.}
	\label{fig_detector}
\end{figure}

\begin{figure}[t]
	\centering
	\includegraphics[width=140mm]{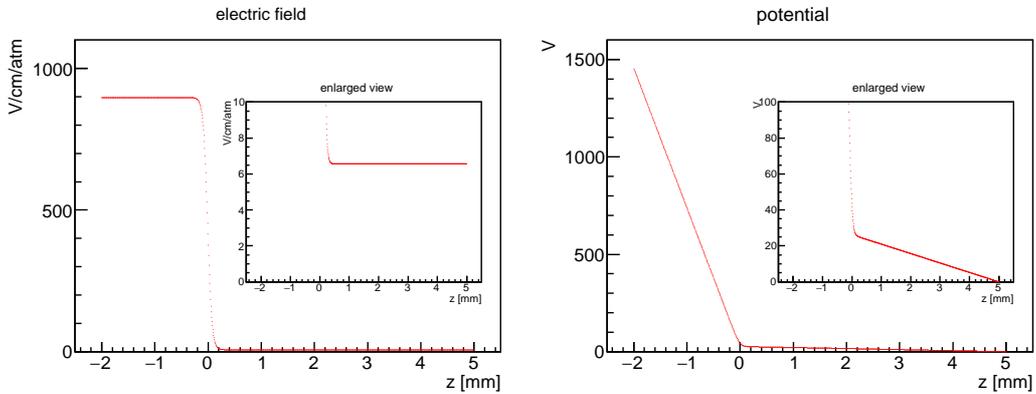}
	\caption{
		Electric field and potential in the detector calculated by gmsh \cite{ref:gmsh} and Elmer \cite{ref:elmer}.
		The drift top, the drift mesh and the EL mesh correspond to z = 5~mm, 0~mm, and -2~mm, respectively.
		0~V, +16~V and +1456~V are applied to the drift top, drift mesh and EL mesh, respectively.
		In the gmsh simulation, the mesh was constructed as wires arranged in x and y directions.
	}
	\label{fig_detector_field}
\end{figure}

\section{Experimental setup}
A detector as depicted in Figure~\ref{fig_detector} was constructed in a vessel which is filled with 8 atm xenon gas during measurements.
The detector has a 5~mm-thick drift region and a 2~mm-thick electroluminescence (EL) region where a high electric field is applied to measure ionization signal via the EL photon detection.
A VUV-sensitive PMT, HAMAMATSU R8520-406, is used for detection of scintillation and EL photons whose wavelength is peaked at around 170~nm.
The effective area of the photocathode of the PMT is 20.5~mm $\times$ 20.5~mm.
Meshes are placed at the bottom of the drift and EL region, and are named drift mesh and EL mesh, respectively.
The drift and EL meshes are woven-type made of gold-plated tungsten wires whose diameter is 0.03~mm and pitch is 100~wires/inch.
We call the cathode of the TPC as the drift top in this paper.
The potential of the drift top is fixed to the ground level, and positive voltages are applied to the meshes to form the electric fields.
To shield the -800~V potential applied to the PMT surface, a ground level mesh is placed 3~mm above the PMT.
Insulators supporting the electrodes are made of PTFE with a 20~mm diameter hole, so as to increase the scintillation yield by reflecting VUV photons.
Ionized electrons move in the drift region toward the EL region, and generate EL photons there.
The z-axis is taken in the direction of the drift electric field, and the direction along the electric field is called vertical, and the direction perpendicular to the electric field is called horizontal in this paper.
Figure~\ref{fig_detector_field} shows the electric field calculated by using gmsh \cite{ref:gmsh} and Elmer \cite{ref:elmer}.
Because of the leakage of the EL field to the drift region, the drift electric field strength is higher than the expectation from the potential difference of the drift mesh and EL mesh: 6.6~V/cm/atm instead of 4~V/cm/atm when +16~V and +1456~V is applied to the drift mesh and EL mesh, respectively (see also the inset of the left panel of Figure~\ref{fig_detector_field}).
We quote values based on the simulation as the drift electric field in this paper.
The effect of the drift electric field on the EL field is less than 0.3\%.
The electron passage rate through the drift mesh is estimated to be $\sim$ 100\% by simulating the drift motion of 100 electrons by using Garfield++ \cite{ref:garfield} when the voltage mentioned above is applied.
\par
A waveform digitizer, CAEN V1720, records signals from the PMT at a 250~MHz sampling rate.
An $^{241}$Am $\alpha$-particle source (5.4~MeV) is set on the cathode electrode.
The thickness of the source is about 3~nm, and the energy loss in the source is negligible.
Actually, $\alpha$-particles with energy differences of about $0.8\%$ of 5485.56 keV at $84.5\%$ and 5442.80 keV at $13.0\%$ are generated from the source.
Since the energy resolution is about $15\%$, the difference in energy of $\alpha$-particles is not considered.
The range of a 5.4~MeV $\alpha$-particle is 2.7~mm in 8~atm xenon gas \cite{ref:SRIM}.
The gas supply and circulation system for this measurement is same as the one in \cite{ref:AXEL2017}.
At the time of measurement, xenon gas was circulated so as not to deteriorate the performance of the detector.
The pump, PumpWorks PW 2070, that can be used in high pressure gas was used, and SAES MicroTorr-MC1-902 and API API-GETTER-I-Re were used in series for purification.

\section{Measurement}\label{sec_mes}
Two types of measurements were performed.
First, we measured the scintillation photons at several drift electric field strengths to confirm that the slow component is made of recombination photons.
Since the recombination process is caused during the electron drift motion, the recombination photons are expected to be detected several $\mu$s after the de-excitation process.
It is also expected the recombination rate is reduced at higher electric field because electrons are more quickly separated from ions.
Second, in order to measure the angular dependence of scintillation, we applied the EL electric field and obtained the event-by-event longitudinal distribution of the ionization electrons from the EL waveform.
When the track is aligned to the electric field direction, more electrons generated at closer position to the EL region, so the EL emission timing is expected to be earlier.

\begin{figure}[t]
	\centering
	\includegraphics[width=140mm]{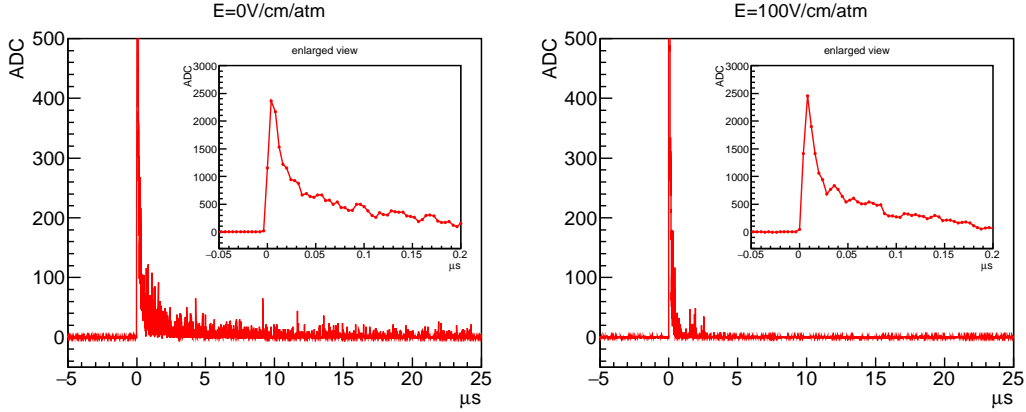}
	\caption{
		Examples of observed waveforms of scintillation signal by a 5.4~MeV $\alpha$-particle.
		The time zero is defined as the rise time of scintillation waveform.
		Left and right figures are those when the drift electric field is 0~V/cm/atm and 100~V/cm/atm, respectively.
	}
	\label{fig_waveform_sci}
\end{figure}
\begin{figure}[t]
	\centering
	\includegraphics[width=80mm]{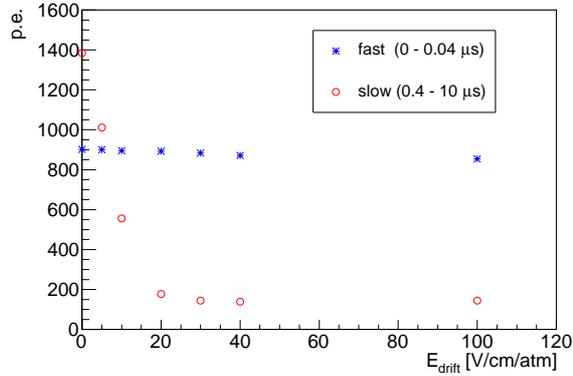}
	\caption{
		The drift electric field dependence of the fast (0-0.04~$\mu$s) and slow (0.4-10~$\mu$s) scintillation components.
	}
	\label{fig_sci-vs-drift}
\end{figure}

\subsection{Electric field dependence of scintillation yield}\label{sec_recombination}
Scintillation yields was measured at several electric field settings.
In this measurement, the EL field was not applied.
Examples of the waveforms at the electric field of 0~V/cm/atm and 100~V/cm/atm are shown in Figure~\ref{fig_waveform_sci}.
One can see more photons after the initial peak when the drift field is not applied.
To quantify this, here, we define the fast and slow components as the photon yield in 0-0.04~$\mu$s and 0.4-10~$\mu$s, respectively. 
Figure~\ref{fig_sci-vs-drift} shows the electric field dependence of the fast and slow components.
It can be seen that when the drift electric field is applied, the slow component decreases.
This observation is consistent with the expectation that more recombination happens and slow component increase at lower electric field.
This phenomenon is reported in \cite{ref:Waseda_NIMA2004,ref:Waseda_NIMA2010}.
Unlike the slow component, the fast component has no dependence of the drift electric field.
The fast component is considered to be de-excitation photons from excited xenon atoms by an $\alpha$-particle mainly with a process: one excited xenon atom collied with an atom and create an exited molecule, which emit one photon \cite{ref:Saito_IEEE2003}.
The slight negative slope of the fast component is thought to be little contribution of the recombination photons.

\begin{figure}[t]
	\centering
	\includegraphics[width=140mm]{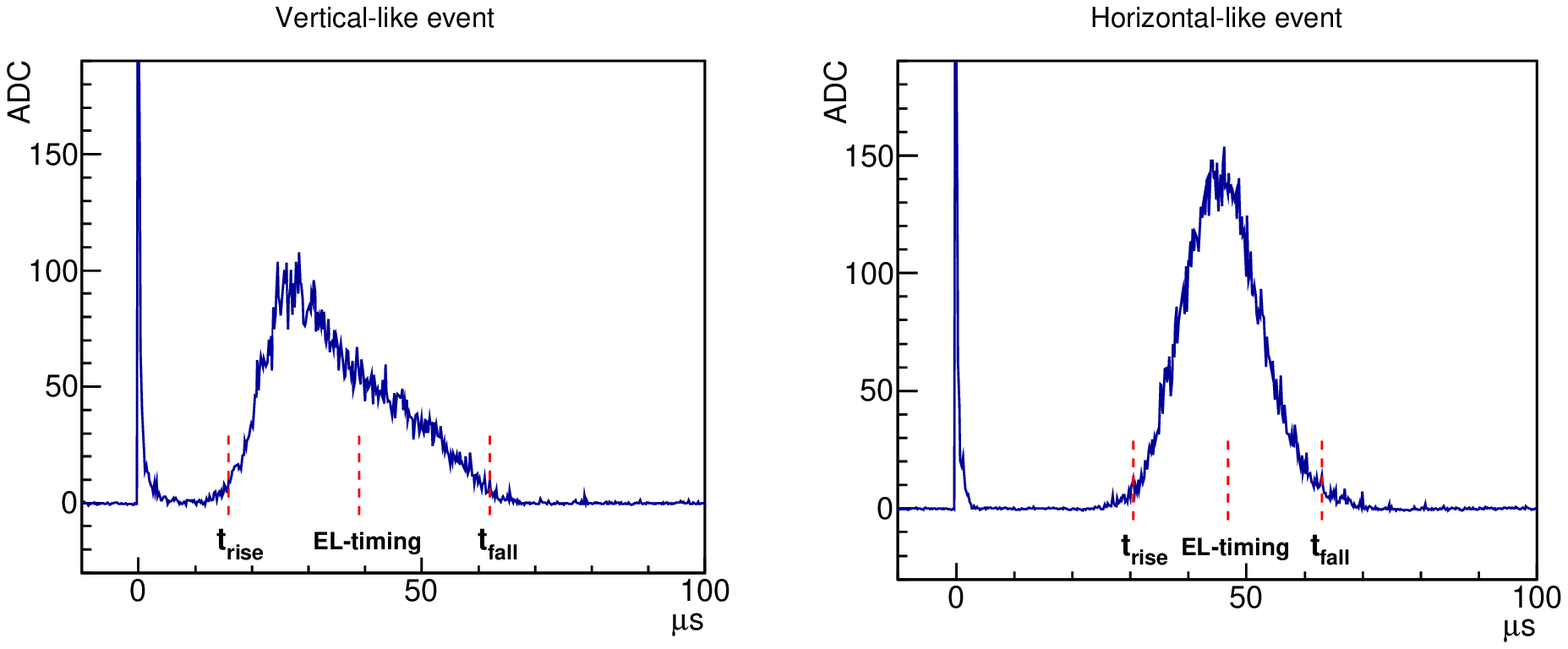}
	\caption{
		Examples of the measured waveform at 6.6~V/cm/atm of the drift electric field and 900~V/cm/atm of the EL field.
		EL photons are observed after the scintillation photons.
		The time profile of the EL waveform represents the ionization electron distribution along the z-axis.
		The left and right figures are typical waveforms when the $\it EL\mathchar`-timing$ parameter is early and late, respectively.
		The red dashed lines show the rise and fall timing.
	}
	\label{fig_waveform_el}
\end{figure}

\begin{figure}[t]
	\centering
	\includegraphics[width=140mm]{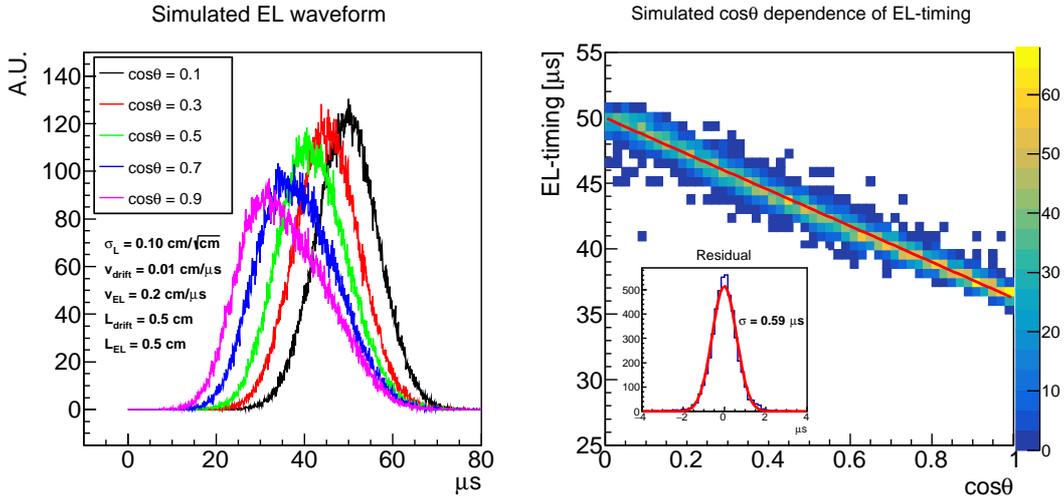}
	\caption{
		Left figure shows the simulated EL waveforms for several cos$\theta$.
		Time 0 is the start timing of the event, which corresponds to the rise time of the scintillation signal in the experimental data.
		Right figure shows the cos$\theta$ dependence of the $\it EL\mathchar`-timing$ parameter by simulation.
		The red straight line is fitted one, and the inset figure shows the residual of the histogram and the line.
		$\theta$ is the angle of $\alpha$-particle direction measured from the drift direction as drawn in Figure~\ref{fig_detector}.		
		In the simulation, the drift velocity of drift (EL) region are set to 0.01~cm/$\mu$s (0.2~cm/$\mu$s) and the longitudinal diffusion coefficient to 0.1~cm/$\sqrt{\rm cm}$.
	}
	\label{fig_sim}
\end{figure}

\subsection{Angular dependence of the yields of scintillation and ionization electrons}
In order to measure the angular dependence of the scintillation yield, we applied the EL electric field.
Figure~\ref{fig_waveform_el} shows waveform examples when 6.6 V/cm/atm and 900~V/cm/atm are applied to the drift and EL regions, respectively.
The ionization electrons drift to the EL region and generate EL photons, hence they are observed after the scintillation photons. 
Since the electrons generated closer to the EL region reach that region earlier, the time profile of the EL photons represents the ionization electron distribution projected to the z-axis.
In the left figure, there are more photons in the earlier time.
This is considered to reflecting the Bragg curve: the $\alpha$-particle is emitted from the drift top and stop at a position closer to the EL field.

We define the parameter $\it EL\mathchar`-timing$ as,
\begin{equation}\label{eq:el_time}
{\it EL\mathchar`-timing}=\frac{t_{\rm rise}+t_{\rm fall}}{2},
\end{equation}
where $t_{\rm rise}$ and $t_{\rm fall}$ are the rise and fall timing of the EL signal, that are determined as follows: rebin the waveform 50 times coarser (corresponding to 5~MHz sampling), set the threshold value at the 8\% of the peak height, find the timings at which the signal crosses the threshold.
The calculated $t_{\rm rise}$, $t_{\rm fall}$ and $\it EL\mathchar`-timing$ are shown in Figure~\ref{fig_waveform_el} as red-dashed lines.
Since the start point of the $\alpha$-particle track is fixed to the drift top, $\it EL\mathchar`-timing$ is expected to depend linearly to cos$\theta$ of the $\alpha$-particle, where $\theta$ is measured from drift direction (-z direction).
Diffusion and the electron distribution would affect this dependence.
We performed a simulation study to estimate this effect.
$\alpha$-particles are generated to get the z distribution of ionization electrons by Geant4\cite{ref:Geant4}
The drift velocity and the longitudinal diffusion constant were tuned so that the distributions of $t_{\rm rise}$ and $t_{\rm fall}$ matches to the observation.
The obtained drift velocity and diffusion constant are 0.01~cm/$\mu$s and 0.1~cm/$\sqrt{\rm cm}$, respectively, and consistent to the measured values in \cite{ref:driftV_PR1962,ref:diffusion1_JJAP2012,ref:diffusion2_JJAP2012}.
The left diagram of Figure~\ref{fig_sim} is EL waveforms by simulation, and the waveform with earlier EL-timing corresponds to a larger cos$\theta$ (vertical direction) event.
The right diagram of　Figure~\ref{fig_sim} shows the correlation between $\it EL\mathchar`-timing$ and cos$\theta$ obtained by the simulation.
The linearity between $\it EL\mathchar`-timing$ and cos$\theta$ is good even with diffusion.
By taking the residual of the histogram and the straight line, the uncertainty was estimated to be 0.59$\mu$s.
Therefore, we use $\it EL\mathchar`-timing$ as a parameter to represent angle.

\begin{figure}[tb]
	\centering
	\includegraphics[width=140mm]{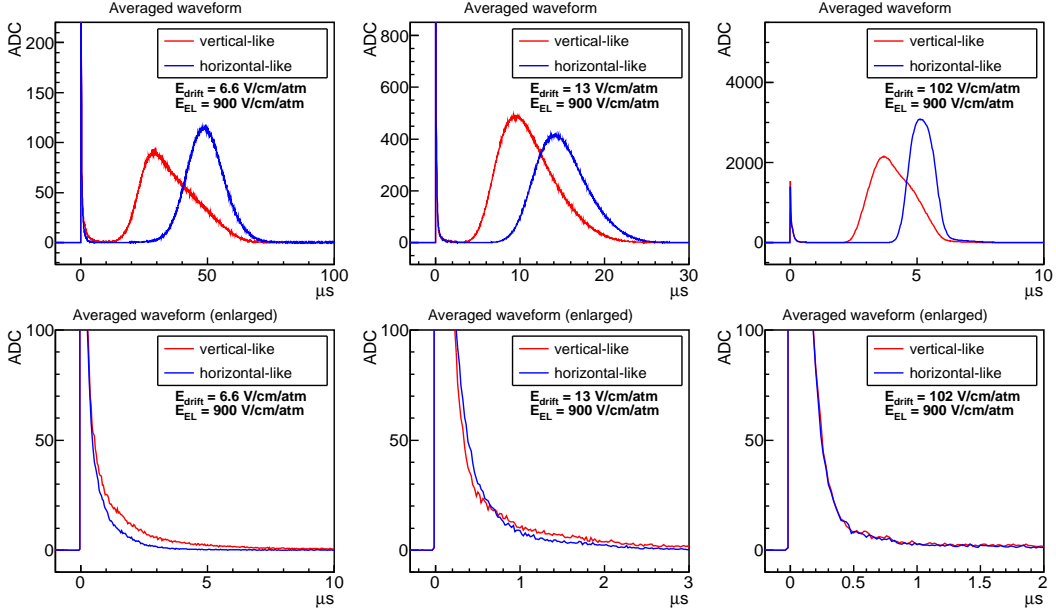}
	\caption{
		Measured averaged waveforms of the scintillation and EL signal for 6.6, 13, and 102~V/cm/atm of the drift electric field.
		Blue (red) lines are the averaged of 20\% events with large (small) $\it EL\mathchar`-timing$, which correspond to the horizontal (vertical) track.
		The plots show waveforms in a wide range so that the EL signal can be seen, and the bottom plots in a narrow range where the scintillation signal is dominant.
	}
	\label{fig_wav_avg}
\end{figure}

Figure~\ref{fig_wav_avg} shows the measured waveforms averaged over many pulses.
Event selections by using $\it EL\mathchar`-timing$ are made to select horizontal or vertical tracks.
Events with $20\%$ from the smaller (larger) $\it EL\mathchar`-timing$ were taken as vertical (horizontal) events.
Concrete values are as shown in Table~\ref{tab_lowE}.
It can be seen that more photons are detected up to 10~$\mu$s for the vertical-like track with 6.6~V/cm/atm of the drift electric field, which is an expected tendency with the columnar recombination.
The timescale of the scintillation emission is consistent with that for ionization electrons overlapping with xenon ions: for the vertical track, it takes 30~$\mu$s for electrons to drift the $\alpha$-particle track length.
At 13~V/cm/atm, similar tendency as for 6.6~V/cm/atm was observed for scintillation signal after 0.6~$\mu$s, but it is not true before 0.6~$\mu$s: the yield is less for vertical-like tracks.
The reason is not known.
The phenomenon itself is interesting, but since understanding is insufficient, data with 13V/cm/atm is not used in analysis to be described later.
At 102~V/cm/atm, there is no difference in scintillation waveforms, and almost all ionization electrons are considered to be drifting.
Based on these observations; the data at the drift electric field of 6.6~V/cm/atm was used for the analysis of columnar recombination.
The data at 102~V/cm/atm is also shown for comparison with the case where columnar recombination does not occur.
As we saw in the section \ref{sec_recombination}, the fast component integration range is set to be 0-0.04~$\mu$s with little recombination contamination.
For 6.6~V/cm/atm (102~V/cm/atm) of the drift electric field, the slow component range is set to 0.4-10~$\mu$s (0.4-2~$\mu$s).
The EL yield is calculated from the integration of 10-100~$\mu$s (2-10~$\mu$s) of the signal.

\begin{figure}[t]
	\centering
	\includegraphics[width=140mm]{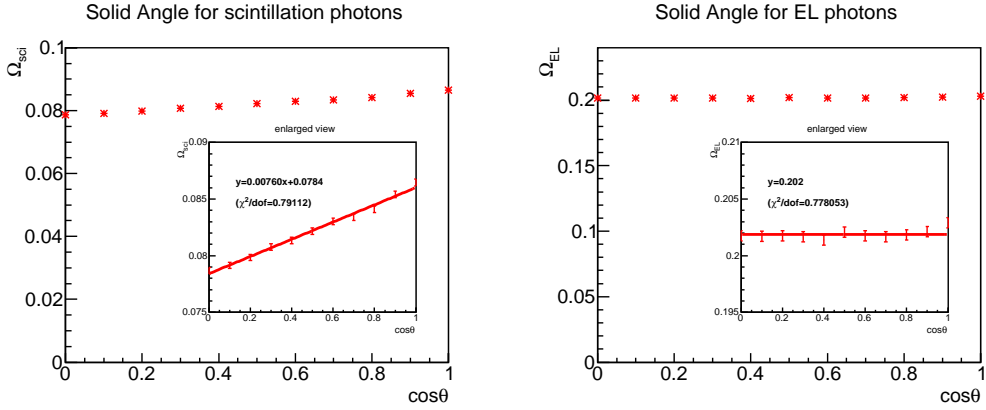}
	\caption{
		Simulated solid angle of the PMT given as a function of cos$\theta$, where $\theta$ is the angle of $\alpha$-particle.
		The reflection on the PTFE insulator (66\%) and drift top (20\%), and the aperture ratio of the meshes (77.8\%) are taken into account.
		The error bars are statistical ones.
		The straight lines are fitted ones with linear function (left) and constant (right).
	}
	\label{fig_solid_angle}
\end{figure}

\begin{figure}[t]
	\centering
	\includegraphics[width=140mm]{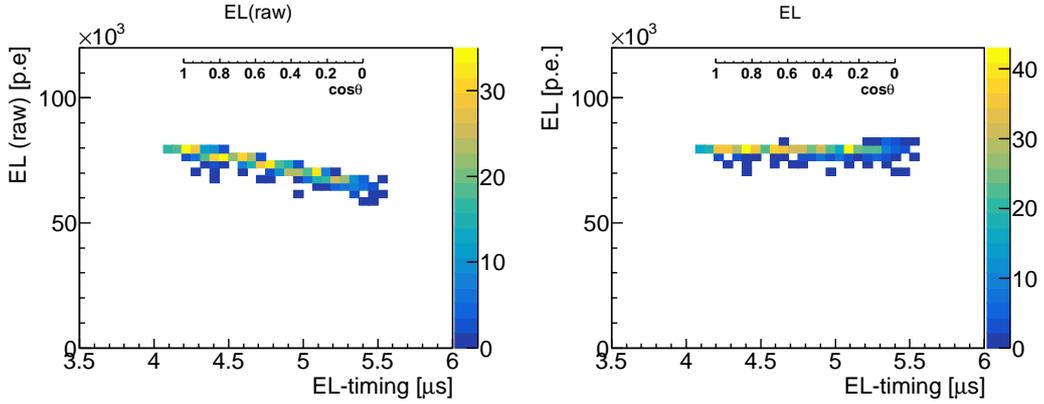}
	\caption{
		Left figure shows the measured $\it EL\mathchar`-timing$ dependence of EL photon yield when 102~V/cm/atm (900~V/cm/atm) is applied as the drift (EL) electric field.
		Corresponding cos$\theta$ values are drawn inside the plot.
		Right figure is same but after the correction.
	}
	\label{fig_hosei}
\end{figure}

\par
The cos$\theta$ dependence of the solid angle of the PMT for scintillation photons ($\Omega_{\rm sci}$) and EL photons ($\Omega_{\rm EL}$) were estimated by simulation and shown in Figure~\ref{fig_solid_angle}.
Since vertical tracks emit their scintillation photons near the PMT, the solid angle is larger compared to horizontal tracks.
The difference is about $10\%$.
We corrected the scintillation yields (fast and slow component) using a linear function obtained with the simulation.
On the other hand, $\Omega_{\rm EL}$ does not depend on cos$\theta$, and we did not correct the EL yield.
\par
The left plot of Figure~\ref{fig_hosei} shows the measured $\it EL\mathchar`-timing$ dependence of the EL yield when 102~V/cm/atm was applied as the drift field, at which recombination is expect to be marginal.
The upper horizontal axes show corresponding cos$\theta$ values.
Though it is expected that there is no angle dependence on the EL yield, the EL yield is smaller for longer $\it EL\mathchar`-timing$ events.
There are two possibilities for $\it EL\mathchar`-timing$ dependence of EL yield.
The first is the attachment, and it can explain that electrons are decreasing as $\it EL\mathchar`-timing$ is longer.
In this case, it is necessary to correct the EL yield in order to know the amount of ionized electrons, and columnar recombination will be enhanced because the EL yield is increased in the horizontal events.
The second case is that ionization electrons collide with the drift plane and are missing.
In this case, not only EL yield but also slow component should be corrected in the same way, and the columnar recombination is enhanced on the EL side, but it is degrade on the slow component side.
Measurement of the attachment with another detector\cite{ref:AXEL2017} using the same gas circulation system as this measurement gave the result that the electron lifetime was 5 ms or more.
This means that the loss of electrons is $1\%$ or less even with drift of 50 $\mu$s.
Since the detector contents are different and the outgas is different, the electron lifetime of this measurement is not necessarily the same, but in this measurement there is a possibility that the attachment has not influenced.
Therefore, the latter model was adopted as a pessimistic condition for columnar recombination, and both EL yield and slow component were corrected by a cos$\theta$ function such that the angular dependence of the EL yield at the high electric field becomes flat as shown in right plot of Figure~\ref{fig_hosei}.

\begin{figure}[t]
	\centering
	\includegraphics[width=150mm]{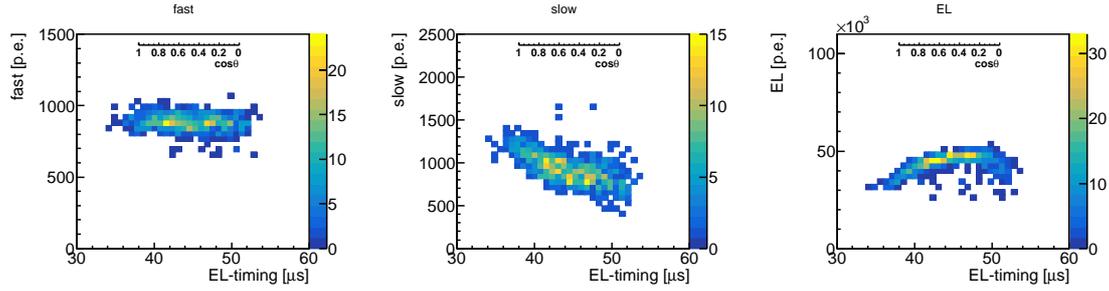}
	\caption{
		The $\it EL\mathchar`-timing$ dependence of the number of photons detected.
		The three plots correspond to the fast component, slow component, and the EL signal from the left.
		The drift electric field was applied at 6.6 V/cm/atm, and the EL one was at 900 V/cm/atm.
	}
	\label{fig_angle_dep_lowE}
\end{figure}
\begin{figure}[t]
	\centering
	\includegraphics[width=150mm]{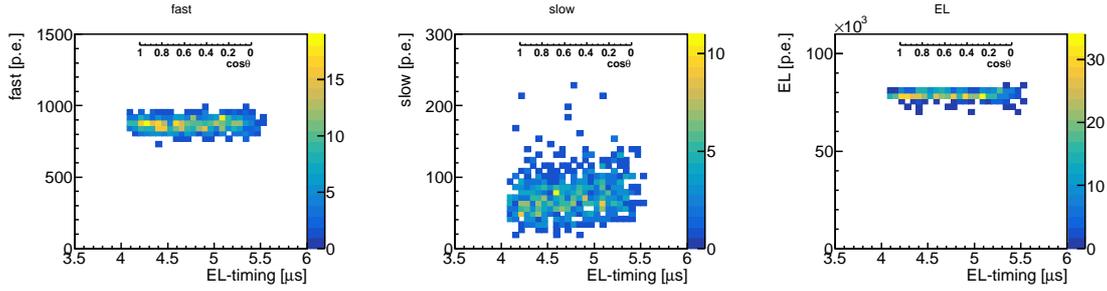}
	\caption{
		Same as Figure~\ref{fig_angle_dep_lowE}, but the 102 V/cm/atm of the drift electric field was applied.
	}
	\label{fig_angle_dep_highE}
\end{figure}

\par
The $\it EL\mathchar`-timing$ dependence of the fast, slow and EL yields measured at the 6.6~V/cm/atm of drift electric field are shown in Figure~\ref{fig_angle_dep_lowE}.
The numerical values are shown in Table~\ref{tab_lowE}.
The fast component, corresponding to the de-excitation of xenon atoms excited by an $\alpha$-particle, shows no dependence of the $\it EL\mathchar`-timing$.
The slow component, which represent the amount of the recombination photon, shows a negative correlation.
The EL yield representing that of the ionization electrons shows a positive correlation.
These correlations indicate the presence of the angular dependent columnar recombination.
This means that the energy and angle can be reconstructed from the scintillation time profile and ionization signal.
The angular resolution obtained from the residual when fitting the correlation between the slow component and the angle $\theta$ with a straight line is 40 degree (FWHM).
\par
Figure~\ref{fig_angle_dep_highE} shows the result with the drift electric field increased to 102 V/cm/atm.
The numerical values are shown in Table~\ref{tab_highE}.
The yield of the fast component is similar to those in Figure~\ref{fig_angle_dep_lowE}, and it is consistent with the fact that the de-excitation process does not depend on the electric field strength.
The slow component is smaller compared to that in Figure~\ref{fig_angle_dep_lowE} and EL yield is larger.
In addition, much weaker dependence on the angle was observed.
This indicates that the columnar recombination is suppressed at higher electric field.
\par

\begin{table}[t]
	\centering
	\smallskip
	\begin{tabular}{|c|c|c|}
		\hline
		& vertical track [p.e.] & horizontal track [p.e.] \\
		& ($\it EL\mathchar`-timing$ < 40.90 $\mu$s) & ($\it EL\mathchar`-timing$ > 48.35 $\mu$s) \\
		\hline
		fast (0-0.04 $\mu$s) &   897 $\pm$   47 &   874 $\pm$   59 \\
		slow (0.4-10 $\mu$s) &  1129 $\pm$  117 &   805 $\pm$  151 \\
		EL (10-100 $\mu$s)   & 38765 $\pm$ 3763 & 44123 $\pm$ 4873 \\
		\hline
	\end{tabular}
	\caption{
		Photon yields when the drift electric field is 6.6~V/cm/atm.
		Events with $20\%$ from the smaller (larger) $\it EL\mathchar`-timing$ were taken as vertical (horizontal) events.
		Concrete values are as shown in the table.
	}
	\label{tab_lowE}
\end{table}
\begin{table}[t]
	\centering

	\smallskip
	\begin{tabular}{|c|c|c|}
		\hline
		& vertical track [p.e.] & horizontal track [p.e.] \\
		& ($\it EL\mathchar`-timing$ < 4.370 $\mu$s) & ($\it EL\mathchar`-timing$ > 5.102 $\mu$s) \\
		\hline
		fast (0-0.04 $\mu$s) &   873 $\pm$  33 &   875 $\pm$   43 \\
		slow (0.4-2 $\mu$s)  &    70 $\pm$  33 &    84 $\pm$   26 \\
		EL (2-10 $\mu$s)     & 79008 $\pm$ 877 & 79197 $\pm$ 1928 \\
		\hline
	\end{tabular}
	\caption{
		Same as Table~\ref{tab_lowE}, but the 102 V/cm/atm of the drift electric field was applied.
	}
	\label{tab_highE}
\end{table}

\section{Conclusion}
We investigated the angular dependence of the columnar recombination in a xenon gas with 5.4 MeV $\alpha$-particles.
Electric field was applied and the time profile of the scintillation signal was used to extract the fast and slow components.
The ionization signal was measured by the electroluminescence (EL) signal.
The track angle $\theta$ was reconstructed from the $\it EL\mathchar`-timing$ and the angular dependence of the signals were measured.
While the fast component shows no dependence, more slow component and less ionization signal were observed for tracks parallel to the electric field.
Our result indicates that both the energy and angle can be reconstructed from the time profile of the scintillation light.
This phenomenon, if established for lower energy nuclear recoil, will open the possibility of the direction-sensitive dark matter search with a large mass detector.

\section*{Acknowledgments}
This work was supported by
JPSP Grant-in-Aid for Scientific Research on Innovative Areas Grant Number 15H01034. 

\bibliographystyle{JHEP}
\providecommand{\href}[2]{#2}\begingroup\raggedright\endgroup

\end{document}